%% file: note.tex
\documentclass{article}

\usepackage{epsfig}

\usepackage{amsmath}
\usepackage{amssymb}
\usepackage{cite}

\usepackage{fancyhdr}
\input{macros}

\providecommand{\href}[2]{#2}

\begin{document}

\input{CS.title.new}

\section{Introduction}

As shown by the authors of ref. \cite{Guralnik:2003we}\footnote{If not
stated otherwise all cross-references of the type (x.y) refer to
formulas in that paper.} the gravitational Chern-Simons term can
be reduced by a Kaluza-Klein like ansatz (3.28), decomposing the 3d metric
into a 2d metric $g_{\mu\nu}$, a $U(1)$ gauge field $A=A_\mu dx^\mu$ and a scalar $\phi$. Invoking conformal invariance, $\phi$ has been set to 1. The resulting 2d action (cf.\
eq.\ (3.35))\footnote{Signs have been adjusted in order to agree with the notation of ref.\ \cite{Grumiller:2002nm}; to compare with ref.\ \cite{Guralnik:2003we} the relations $R=-r$ and $F=f$ are helpful.}
\eq{
L\left[g_{\mu\nu},A_\mu\right]=\frac{1}{8\pi^2}\int_{\mathcal{M}_2} d^2x\sqrt{-g}\left(FR-F^3\right)\,,
}{cs:0}
depends on the 2d curvature scalar $R$ and on the abelian dual field strength $F=-2\ast d A$. It thus
represents a 2d field theory of gravity interacting with the gauge
field 1-form $A$. 

Classical solutions have been constructed {\em locally} in ref.\ \cite{Guralnik:2003we}, labelled by a constant of motion $c$ whereby another constant of motion has been fixed to a certain value. As far as curvature is concerned this discussion has been exhaustive; however, as will be shown in this work, isocurvature solutions exists with a different number (and different types) of Killing horizons. The main purpose of this note is to elevate the discussion to a {\em global} level, i.e.\ to construct all possible Carter-Penrose (CP) diagrams. A condensed version of the results is plotted in fig.\ \ref{fig:1}. 

An action like (\ref{cs:0}) is equivalent to a first
order gravity action which, in turn, is a special case of a  
Poisson-sigma model (PSM) \cite{Ikeda:1993aj} 
\eq{
L=\frac{1}{4\pi^2}\int_{\mathcal{M}_2} \left[X_a (D\wedge e)^a +Xd\wedge\omega + Y d\wedge A + \epsilon \mathcal{V} (X,Y) \right]\,,
}{cs:1}
with target space coordinates $Y,X,X^+,X^-$, gauge field 1-forms $A,\om,e^-,e^+$ and
\eq{
\mathcal{V} (X,Y) = \frac 12 \left(XY-X^3\right)\,.
}{cs:2}
The notation of ref.\ \cite{Grumiller:2002nm} is used.\footnote{$e^a$ is the
zweibein one-form, $\epsilon = e^+\wedge e^-$ is the volume two-form. The one-form
$\omega$ represents the  spin-connection $\om^a{}_b=\eps^a{}_b\om$
with  the totally antisymmetric Levi-Civit{\'a} symbol $\eps_{ab}$ ($\eps_{01}=+1$). With the
flat metric $\eta_{ab}$ in light-cone coordinates
($\eta_{+-}=1=\eta_{-+}$, $\eta_{++}=0=\eta_{--}$) the first
(``torsion'') term of (\ref{cs:1}) is given by $X_a(D\wedge e)^a =
\eta_{ab}X^b(D\wedge e)^a =X^+(d-\omega)\wedge e^- +
X^-(d+\omega)\wedge e^+$. Signs and factors of the Hodge-$\ast$ operation are defined by $\ast\epsilon=1$. The target space coordinates $X,X^a$ can be interpreted as Lagrange multipliers for geometric curvature and torsion, respectively.} In addition to the Cartan variables an
abelian gauge field 1-form $A$ is present and a new target space
coordinate $Y$ which acts as Lagrange multiplier for gauge curvature. Theories of that type are known for a long time
\cite{Kummer:1995qv}. Actually the transition from (\ref{cs:1}) to (\ref{cs:0}) is
very easy. Variation of $Y$ in (\ref{cs:1}) yields $X=-2\ast dA=F$ where $F$ is
precisely the dual field strength in (\ref{cs:0}). Because this equation is linear in $X$,
the re-insertion of $X$ into the variational principle is permitted. 
A similar argument allows the replacement of the the spin connection
by its dependent form (cf. (\ref{eq:a8}) below) because the first term of (\ref{cs:1}) requires
vanishing torsion. The second term with the dependent spin connection
in $R = 2\ast d\om$ immediately leads to the first term of (\ref{cs:0}).
In (\ref{cs:1}) the corresponding Poisson tensor has rank 2, apart from one
point in target space, namely $X^a = X = Y = 0$. Therefore, the number of Casimir functions is two (in physical terms they correspond to the conserved total charge $c$  and energy $\mathcal{C}^{(g)}$).
A reformulation (\ref{cs:1}) is advantageous because 
powerful tools exist to deal with PSMs at the classical and quantum
level \cite{Ikeda:1993aj,Cattaneo:1999fm}. %\footnote{That such a relation could exist can be suspected on general grounds, because Chern-Simons theory is related to a WZW model on the boundary \cite{Gawedzki:2001ye} which in its gauged version can be reformulated as PSM \cite{Alekseev:1995py}), but it can also be shown explicitly in a few lines.}  
Further details on first order
gravity and a more comprehensive list of references can be found in ref.\
\cite{Grumiller:2002nm}.

With eq.\ (\ref{cs:1}) as the starting point all classical solutions can be
determined with ease. The  solution discussed in ref.\
\cite{Guralnik:2003we} is found to be a special case where the
Casimir functions $c$ and $\mathcal{C}^{(g)}$ are related in a special manner. In general each solution is labelled by the constant values of $c$
and $\mathcal{C}^{(g)}$ and is valid in a certain patch of coordinates. From those
patches all global solutions can be found the structure of which is summarized in fig.\ \ref{fig:1}. Finally, possible generalizations are pointed out.

\section{All classical solutions}

The equations of motion (EOM) for the action (\ref{cs:1}) read:
\begin{eqnarray}
 &  & dY = 0\,, \label{cs:3} \\
 &  & dX+X^{-}e^{+}-X^{+}e^{-}=0\, ,\label{eq:a5} \\
 &  & (d\pm \omega )X^{\pm }\pm \frac 12 (X^3-XY)e^{\pm }=0\, ,\label{eq:a6} \\
 &  & dA+\epsilon \frac 12 X =0\,,\label{cs:4} \\
 &  & d\omega -\epsilon \frac 12 (3X^2-Y)=0\, ,\label{eq:a7} \\
 &  & (d\pm \omega )e^{\pm }=0\, .\label{eq:a8}  
\end{eqnarray}
The action (\ref{cs:1}) is mapped into $-L$ by the $\mathbb{Z}_2$
transformation  $X\to -X$, $X^\pm\to -X^\pm$, $A\to -A$.\footnote{In the second order approach the same discrete symmetry has been observed in ref.\ \cite{Guralnik:2003we} (cf.\ the comment below (4.49)).} An important
distinguishing feature as compared to dilaton gravity coupled to an
abelian gauge field  is the  term $XY$ present in (\ref{cs:2}) because it
is linear in $Y$. By contrast a typical abelian gauge theory with $F^2$
in the second order form would require a term proportional to $Y^2$ in (\ref{cs:2}), as can
be checked easily.

The integration of (\ref{cs:3}) immediately yields the first Casimir
function, $Y=c=\rm const.$ which may be interpreted as ``charge''. The second, geometric one (cf.\ e.g.\ (3.23) of ref.\ \cite{Grumiller:2002nm}), the ``energy'', is obtained by multiplying eqs.\ (\ref{eq:a6})
respectively by $X^-,X^+$, adding them and inserting (\ref{eq:a5}): 
\eq{
\mathcal{C}^{(g)}=X^+X^--\frac 18 X^4+\frac Y4 X^2\,.
}{cs:5}
Eq.\ (\ref{cs:4}) implies $X=-2\ast dA$, thus the dual field strength $F$ is determined by the ``dilaton'' field $X$. The last equation (\ref{eq:a8}) entails the condition of vanishing torsion and can be used to solve for the spin-connection $\om=\eta_{ab}e^a\ast d\wedge e^b$.

\subsection{Constant dilaton vacua}

For $X^+=0=X^-$ eq.\ (\ref{eq:a5}) implies $X=\rm const.$ From (\ref{eq:a6}) it can be deduced immediately that only three solutions are possible: a $\mathbb{Z}_2$ symmetric one ($X=0$) and two non-symmetric ones ($X=\pm\sqrt{c}$, $c>0$). The solutions for the curvature scalar $R=-c$ resp.\ $R=2c$ from (\ref{eq:a7}) indicate (A)dS space (cf.\ (4.50) and (4.51)). The corresponding line element can be presented as\footnote{In fact such solutions exist if $X^+=0=X^-$ in generic 2d gravity theories (\ref{cs:1}) when a more general potential $\tilde{\mathcal{V}}(X^+X^-,X,Y)$ permits one or more solutions to the algebraic equation $\tilde{\mathcal{V}}(0,X,c)=0$. There are as many distinct vacua as there are solutions to that equation. Curvature is given by $R=-2\partial\tilde{\mathcal{V}}/\partial X$. Even if $\tilde{\mathcal{V}}$ depends on $X^+X^-$, $\om$ remains the Levi-Civit{\'a} connection.}
\eq{
(ds)^2=2dudx+\left(\frac{R}{2}x^2+Ax+B\right)(du)^2\,,
}{cs:23}
with some integration constants $A,B$ which have to be fixed appropriately. They are neither defined by the first Casimir $c$ (which enters $R$) nor by the second one $\mathcal{C}^{(g)}$. The latter vanishes for the symmetric solution and becomes equal to $\mathcal{C}^{(g)}=c^2/8$ for the non-symmetric ones. The global structure is the same as the one of the Jackiw-Teitelboim (JT) model \cite{Barbashov:1979bm}, namely (A)dS space.

\subsection{Generic solutions}

All other classical solutions can be constructed in the usual manner \cite{Kummer:1995qv,Klosch:1996fi}. In a patch where $X^+\neq 0$ one obtains\footnote{If $X^+=0$ and $X^-\neq 0$ then the same procedure can be applied with $+\leftrightarrow -$. If both $X^+=0=X^-$ in an open region we have the constant dilaton vacuum discussed above.} the line element in Eddington-Finkelstein (EF) gauge 
\eq{
(ds)^2=2dudX+K(X;\mathcal{C}^{(g)},c)(du)^2\,,\quad K(X;\mathcal{C}^{(g)},c)=2\mathcal{C}^{(g)}-\frac c2 X^2+\frac 14 X^4\,.
}{cs:6}
Evidently there is always a Killing vector\footnote{This is a general feature of $2D$ first order gravity actions (\ref{cs:1}), albeit it is {\em not} a feature of generic $2D$ gravity. This property was also noted in appendix A of ref.\ \cite{Guralnik:2003we}.} $k^\al\partial_\al=\partial/\partial u$ with associated Killing norm $g_{\al\be}k^\al k^\be=K(X;\mathcal{C}^{(g)},c)$. The curvature scalar becomes
\eq{
R = d^2K/dX^2 = -c+3X^2\,. 
}{cs:7}
Obviously, solutions with constant curvature are only possible for the constant dilaton vacuum. With the coordinate redefinition (cf.\ eq.\ (4.52))
\eq{
X=:\sqrt{c}\tanh{y}\,,\quad y:= \left(\frac{\sqrt{c}}{2}z\right)\,,
}{cs:8}
curvature transforms to
 \eq{
R = -c+3c\tanh^2 y = 2c - \frac{3c}{\cosh^2{y}}\,. 
}{cs:9}
This is consistent with (4.53). With the Ansatz $du=\al dt+\be(z)dz$ and (\ref{cs:8}) the line element (\ref{cs:6}) can be brought into diagonal form:
\eq{
(ds)^2=\frac{1}{\cosh^4{y}}\left(1+\de\right)(dt)^2-\frac{(dz)^2}{1+\de}\,, \quad \de:=\left(8\mathcal{C}^{(g)}/c^2-1\right)\cosh^4{y}\,.
}{cs:10}
In the special case $c^2=8\mathcal{C}^{(g)}$ eq.\ (\ref{cs:10}) coincides with eq.\ (4.54). % In that case the Killing norm in (\ref{cs:6}) coincides with the potential $V(f)$ in eq.\ (4.57).

%Obviously the coordinate transformation (\ref{cs:8}) becomes singular for $X=\pm\sqrt{c}$. 
Whenever a diagonal gauge of this type is chosen for a geometry exhibiting Killing horizons coordinate singularities appear. As a consequence the line element (\ref{cs:10}) acquires a coordinate singularity at $X=\pm\sqrt{c}$. Therefore, the line element in EF gauge (\ref{cs:6}) is a more suitable starting point for a discussion of the global structure because it allows for an extension across Killing horizons.

\section{Global properties}

Applying well-known methods \cite{Walker:1970,Klosch:1996qv} the first step of a global discussion is to construct the building blocks of the CP diagrams. The second step is to find their consistent geodesic extensions. In a third step solutions of more complicated topology can be arranged \cite{Klosch:1997md}. Finally, one can try to identify patches in a nontrivial way in order to obtain kink solutions \cite{Klosch:1998yh}.

\subsection{Building blocks}

The basic patches are represented by CP diagrams derived from the metric in EF form (\ref{cs:6}), together with their mirror images (the flip corresponds essentially to a change from ingoing to outgoing EF gauge or vice versa). They determine the set of building blocks from which the global CP diagram is found in a next step by geodesic extension.

The Killing norm $K$ in (\ref{cs:6}) has the form of a Higgs potential. Its four zeros are given by
\eq{
X_h^{1,2,3,4} = \pm\sqrt{c\pm\sqrt{c^2-8\mathcal{C}^{(g)}}}\,.
}{cs:14}
Only for real zeros a Killing horizon emerges. There are several possibilities regarding the number and type of Killing horizons. For positive $c$ any number from 0 to 4 is possible, for negative or vanishing $c$ just 0, 1 or 2 horizons can arise. In all CP diagrams bold lines correspond to the curvature singularities encountered at $X\to\pm\infty$. Dashed lines are Killing horizons (multiply dashed lines are extremal ones). The lines of constant $X$ are depicted as ordinary lines. The triangular shape of the outermost patches is a consequence of the asymptotic behavior ($X\to\pm\infty$) of the Killing norm. The singularities are null complete (because $X$ diverges) but incomplete with respect to non-null geodesics, because the ``proper time'' (cf.\ eq.\ (3.50) of ref.\ \cite{Grumiller:2002nm}; $A=\rm const.$)
\eq{
\tau=\int^XdX'/\sqrt{|A-K(X')|} = {\rm const.} - {\cal O}\left(\frac{1}{X}\right)\,,
}{cs:1783}
does not diverge at the boundary. This somewhat counter intuitive feature has been witnessed already for the dilaton black hole \cite{Katanaev:1997ni}. Regarding this property the singularities differ essentially from the ones in the JT model which are complete with respect to all geodesics.

``Time'' and ``space'' in conformal coordinates should be plotted in the vertical resp.\ horizontal direction. Therefore, all diagrams below except {\bf B0} should be considered rotated clockwise by $45^o$.

\paragraph{No horizons}

If $K$ has no zeros no Killing horizons arise. This happens for positive $c$ provided that $8\mathcal{C}^{(g)}>c^2$ and for $c\leq 0$ if $\mathcal{C}^{(g)}>0$. Modulo completeness properties this diagram is equivalent to the one of the JT model when no horizons are present (cf.\ e.g.\ fig.\ 9 in ref.\ \cite{Klosch:1996qv}).

\vspace{0.1cm}

\begin{minipage}[c]{0.1\linewidth}
{\bf B0:}
\end{minipage}
\begin{minipage}[c]{0.3\linewidth}
\epsfig{file=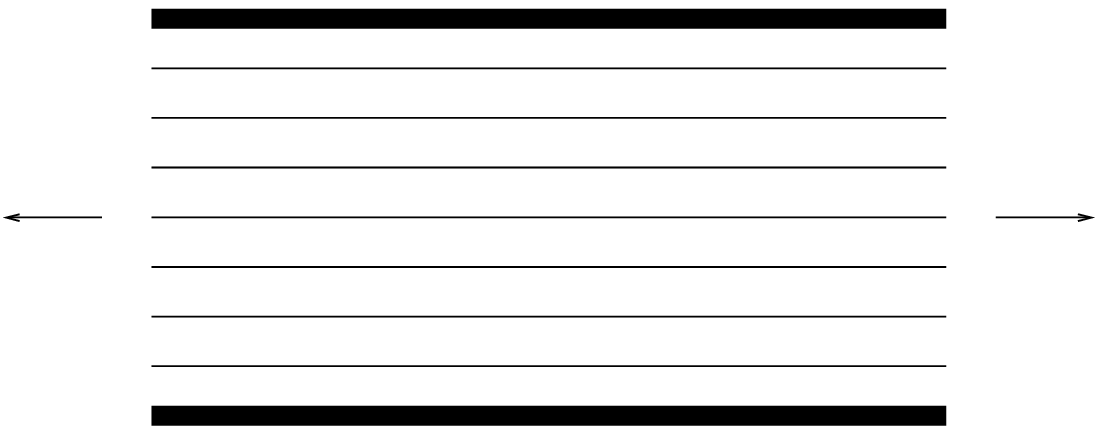,width=\linewidth}
\end{minipage}

\paragraph{One extremal horizon}

This scenario can only happen for $c\leq 0$ (if the inequality is saturated the zero in the Killing norm is of fourth order, otherwise just second order). Additionally, $\mathcal{C}^{(g)}$ must vanish. The horizon is located at $X=0$.

\vspace{0.1cm}

\begin{minipage}[c]{0.1\linewidth}
{\bf B1a:}
\end{minipage}
\begin{minipage}[c]{0.25\linewidth}
\epsfig{file=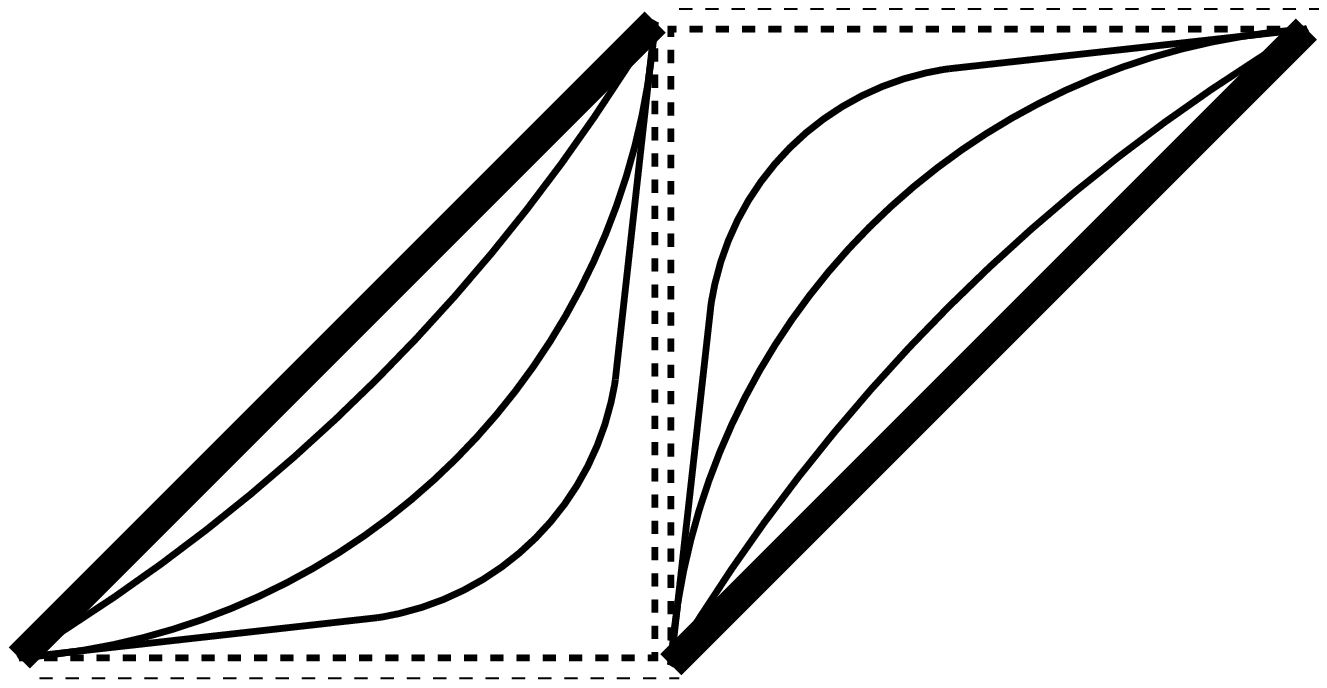,width=\linewidth}
\end{minipage}
\begin{minipage}[c]{0.1\linewidth}
\,
\end{minipage}
\begin{minipage}[c]{0.1\linewidth}
{\bf B1b:}
\end{minipage}
\begin{minipage}[c]{0.25\linewidth}
\epsfig{file=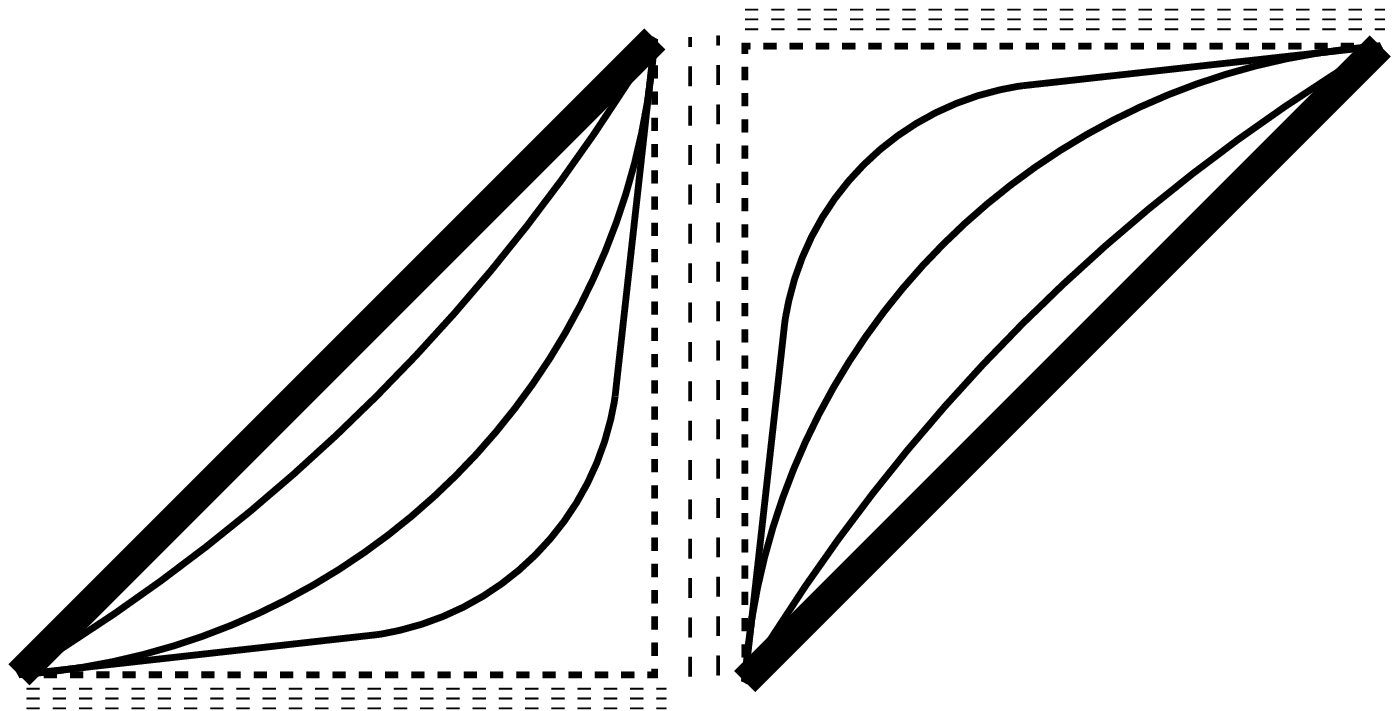,width=\linewidth}
\end{minipage}

\paragraph{Two horizons}

For negative $\mathcal{C}^{(g)}$ and arbitrary $c$ two horizons arise at $X=\pm\sqrt{c+\sqrt{c^2-8\mathcal{C}^{(g)}}}$. Modulo completeness properties this diagram is equivalent to the one of the JT model when two horizons are present.

\vspace{0.1cm}

\begin{minipage}[c]{0.1\linewidth}
{\bf B2a:}
\end{minipage}
\begin{minipage}[c]{0.35\linewidth}
\epsfig{file=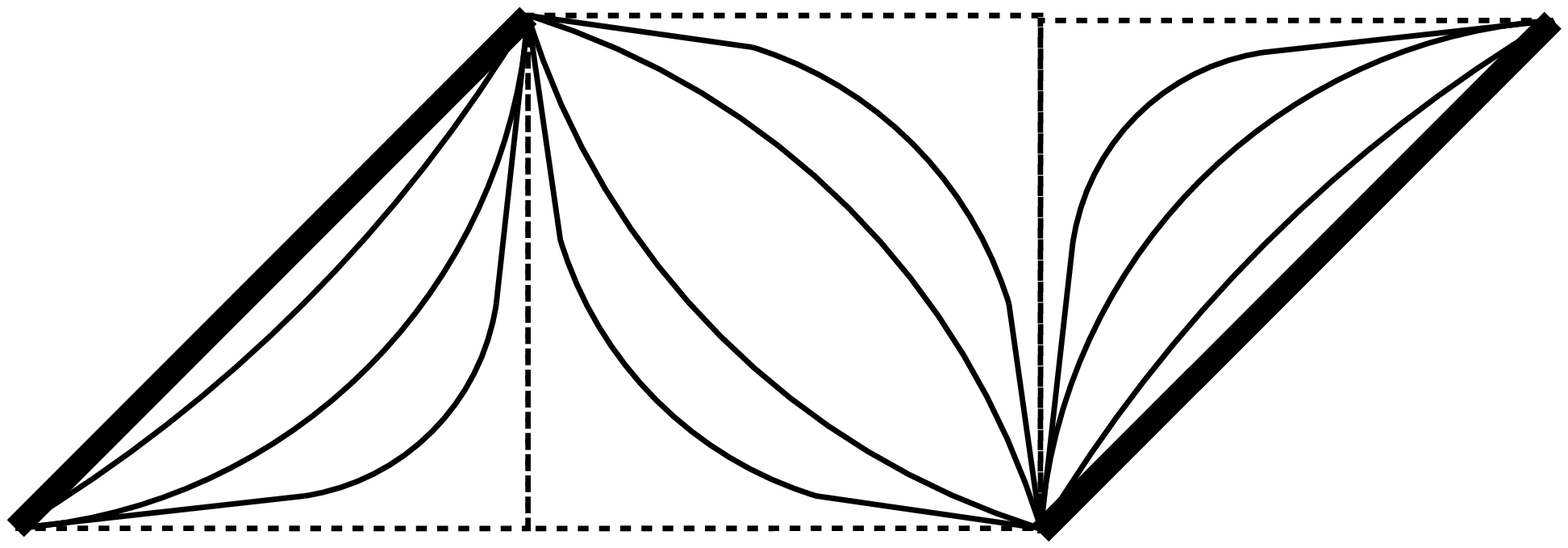,width=\linewidth}
\end{minipage}

\paragraph{Two extremal horizons}

This special case appears for $c>0$ and $c^2=8\mathcal{C}^{(g)}$. The square patch in the middle corresponds to the nontrivial solution discussed in ref.\ \cite{Guralnik:2003we}.

\vspace{0.1cm}

\begin{minipage}[c]{0.1\linewidth}
{\bf B2b:}
\end{minipage}
\begin{minipage}[c]{0.35\linewidth}
\epsfig{file=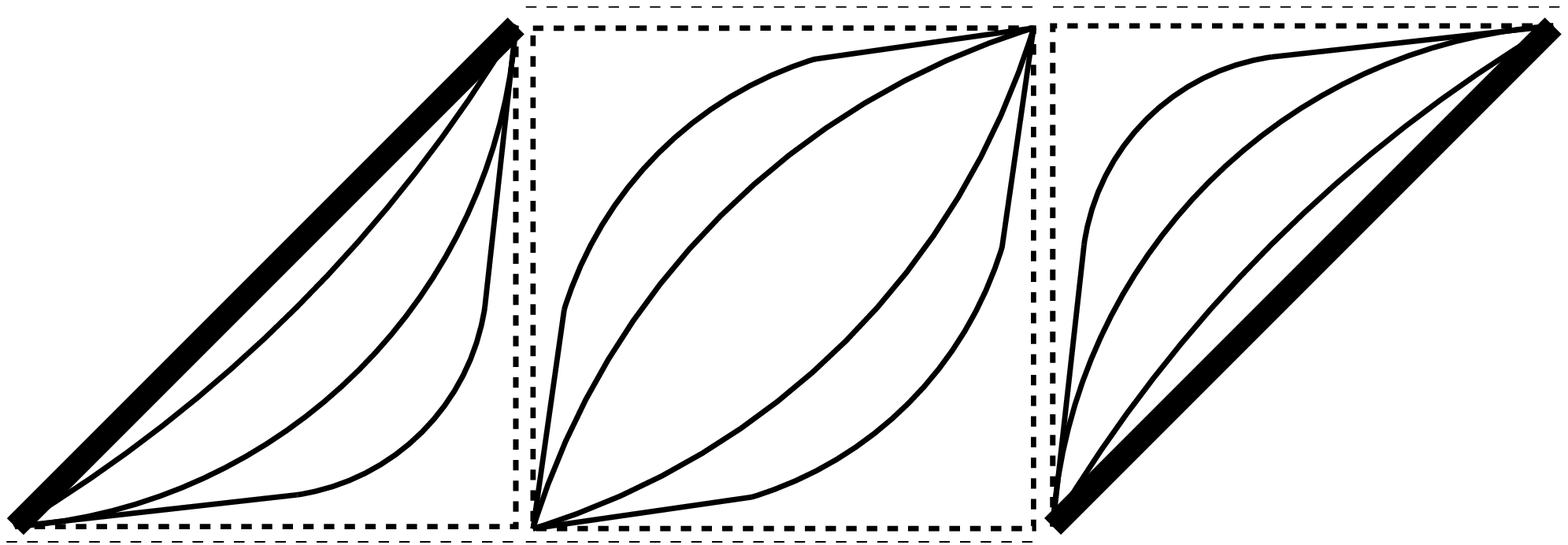,width=\linewidth}
\end{minipage}

\paragraph{Two horizons and one extremal horizon}

If $c>0$ and $\mathcal{C}^{(g)}=0$ an extremal horizon at $X=0$ is present. The two non-extremal ones are located at $X=\pm\sqrt{2c}$. This building block will generate non-smooth CP diagrams due to the appearance of extremal {\em and} non-extremal horizons (cf.\ fig.\ 3 of ref.\ \cite{Klosch:1996qv} and the discussion on that page).

\vspace{0.1cm}

\begin{minipage}[c]{0.1\linewidth}
{\bf B3:}
\end{minipage}
\begin{minipage}[c]{0.5\linewidth}
\epsfig{file=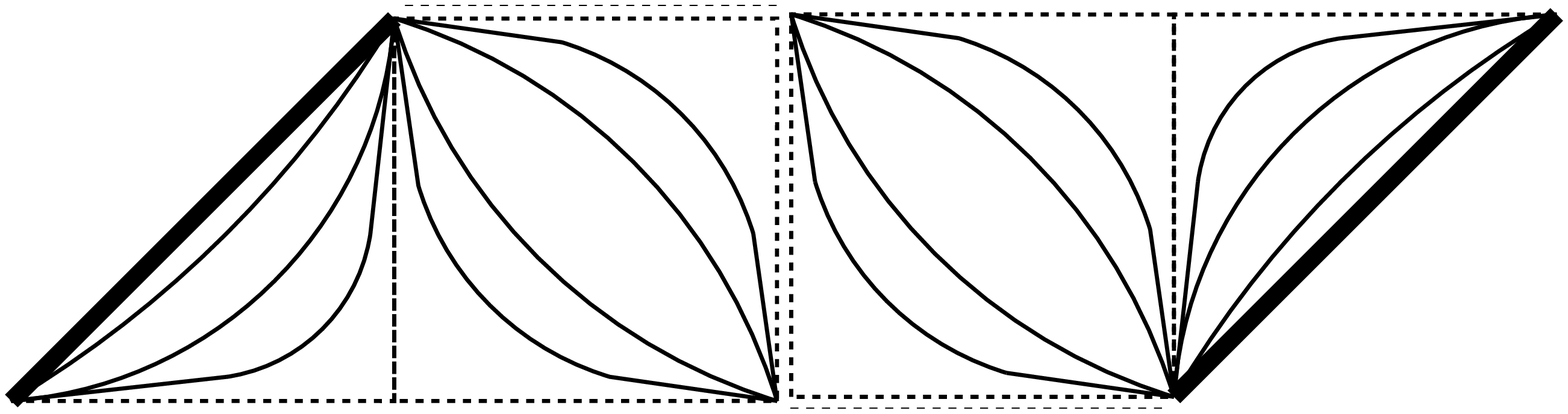,width=\linewidth}
\end{minipage}

\paragraph{Four horizons}

For $c>0$ and $c^2>8\mathcal{C}^{(g)}>0$ four horizons are present given by (\ref{cs:14}).

\vspace{0.1cm}

\begin{minipage}[c]{0.1\linewidth}
{\bf B4:}
\end{minipage}
\begin{minipage}[c]{0.6\linewidth}
\epsfig{file=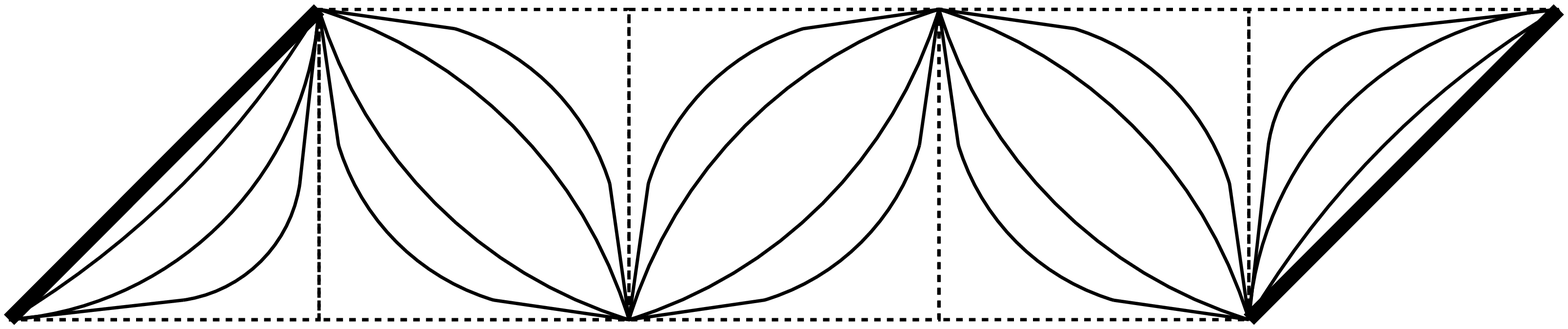,width=\linewidth}
\end{minipage}

\subsection{Maximal extensions}

The boundaries of each building block are either geodesically complete (infinite affine parameter with respect to all geodesics) or incomplete otherwise. Loosely speaking, when in the latter case a curvature singularity is encountered no continuation is possible. For an incomplete boundary without such an obstruction appropriate gluing of patches provides a geodesic extension. Identifying overlapping squares and triangles of each type of building block in this manner the full CP diagram is constructed. Generically basic patches with 3 or more horizons produce 2d webs rather than onedimensional ribbons as global CP diagrams. Here, as a nontrivial consequence of the triangular shape at {\em both} ends of the building blocks, with the diagonal oriented in the {\em same} direction, the allowed topologies drastically simplify to a ribbon-like structure.\footnote{Such a structure is rather typical for theories with charge and mass. The most prominent example is the Reissner-Nordstr{\"om} black hole.} 

\begin{figure}[htb]
\centering
\epsfig{file=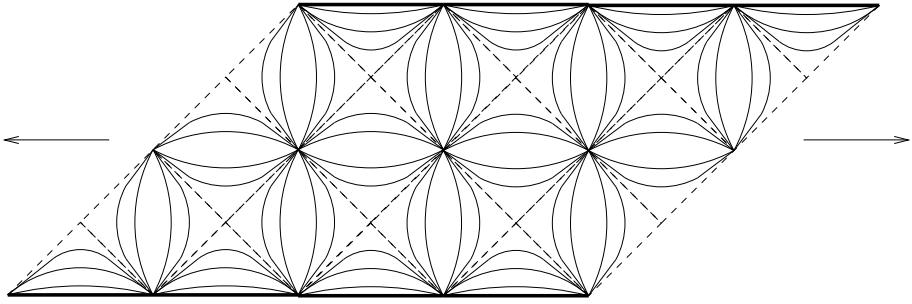,width=\linewidth}
\caption{Maximally extended CP diagram for the four horizon scenario}
\label{fig:ribbon}
\end{figure}

{\bf B0} already coincides with its maximal extension. The one of {\bf B4} is depicted in fig.\ \ref{fig:ribbon}. 
All other global diagrams with a smaller number of horizons can be obtained from this one by contracting appropriate patches and by adding dashed lines if extremal horizons are present. For instance, the one horizon cases {\bf B1a} and {\bf B1b} can be obtained by eliminating all square patches and adding either one or three dashed lines. 

There are up to three types of vertex points in these diagrams: vertices between the singularities along the border, vertices where lines $X=\rm const.$ from 4 adjacent patches meet (``sources'' or ``sinks'' for Killing fields) and vertices which are similar to the bifurcation 2-sphere of Schwarzschild spacetime.  Their (in)completeness properties follow from (\ref{cs:1783}) for $A=0$ (so-called ``special geodesics''): $\tau=\int^XdX'|K(X')|^{-1/2}$. Thus, the vertices at the boundary are incomplete. All other vertices are incomplete if no extremal horizon is present, because (\ref{cs:1783}) remains finite for $A=0$ only at nondegenerate horizons. 

Of course, as in the Reissner-Nordstr{\"o}m case, one can identify periodically (e.g.\ by gluing together the left hand side with the right hand side in fig.\ \ref{fig:ribbon}). M{\"o}bius-strip like identifications are possible as well.

\subsection{Kinks}
%\subsection{The Kink is Dead -- Long Live the Kink!}

From a global point of view the ``kink'' solution discussed in ref.\ \cite{Guralnik:2003we} consists of the two symmetry breaking constant dilaton vacuum solutions in the regions $|X|>\sqrt{c}$ and the square patch of {\bf B2b} inbetween.

Such a patching in general induces a matter shock wave at the connecting boundary. For $C^1$ solutions no patching of that kind is possible in the framework of PSMs \cite{Bojowald:2003pz} simply because either $X^+$ or $X^-$ becomes discontinuous (in one region it is non-vanishing, in the others it is identical to zero). 

It is illustrative to discuss in more detail what happens if one joins (\ref{cs:23}) to (\ref{cs:6}). By adjusting $A$ and $B$ the Killing norm can be made $C^2$. Hence curvature becomes continuous. Nevertheless, the discontinuity of $X^+$ in eq.\ (\ref{eq:a6}) implies the existence of matter at the horizon (the version of (\ref{eq:a6}) with matter is given by eq.\ (3.8) of ref.\ \cite{Grumiller:2002nm}) with a localized energy-momentum 1-form
\eq{
T^+:=\frac{\de L^{(m)}}{\de e^-} = \left(\de(x-\sqrt{c})-\de(x+\sqrt{c})\right)dx\,, \quad T^-:=\frac{\de L^{(m)}}{\de e^+} = 0\,,
}{cs:loc}
where $L^{(m)}$ is the induced matter action. The coordinate $x$ is the same as in (\ref{cs:23}). It coincides with $X$ for $X^2\leq c$.

This problem is not evident if the coordinate system (\ref{cs:10}) is used because the matter sources are pushed to $z=\pm\infty$. But patching at a coordinate singularity like the one at these points is difficult to interpret. It is therefore not quite clear in what sense the solution presented in ref.\ \cite{Guralnik:2003we} can be considered as kink from a global point of view.

Actually general methods exist which allow the construction of kink solutions taking the global diagrams as a starting point \cite{Klosch:1997md,Klosch:1998yh}. As noted above the ribbon-like CP diagrams related to {\bf B0}-{\bf B4} allow for periodic identifications. If they are performed in a nontrivial manner as in fig.\ 9 of ref.\ \cite{Klosch:1998yh} this provides one way kink solutions may appear. It could be rewarding to study them at the level of $2D$ dilaton gravity in the first order formulation in order to learn more about non-trivial sectors of Chern-Simons theory in $3D$.

\section{Outlook}

\begin{figure}
\centering
\epsfig{file=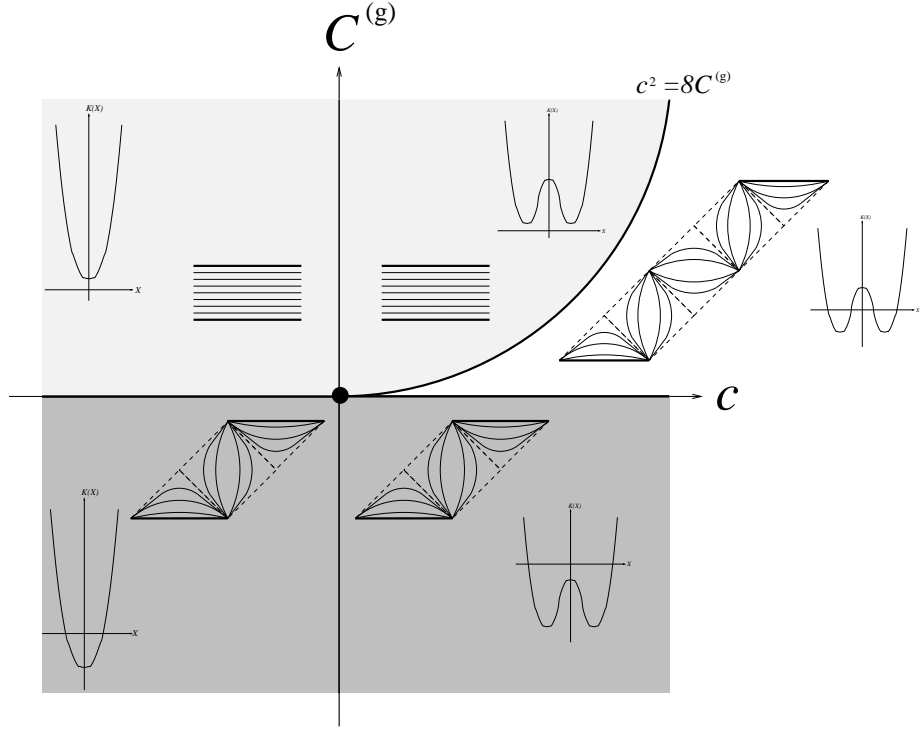,width=\linewidth}
\caption{The phase space of building blocks for general CP diagrams. The white, dark gray and light gray region contains all CP diagrams with four, two and zero non-extremal Killing horizons, respectively. Bold lines in the phase diagram correspond to CP diagrams containing one or two extremal horizons (and possibly additional non-extremal ones). The point at the center corresponds to the special case $c=0=\mathcal{C}^{(g)}$ with an {\em extremely} extremal horizon (with fourth order zero in the Killing norm). The solution found in ref.\ \cite{Guralnik:2003we} corresponds to the curved bold line separating the white from the light gray region. In the CP diagrams bold, dashed and ordinary lines correspond to curvature singularities, non-extremal Killing horizons and $X=\rm const.$ lines, respectively (only the non-extremal cases are depicted). The Killing norm as a function of $X$ also has been plotted in the five non-extremal regions (in the extremal limit zeros can be located at some of the extrema).}
\label{fig:1}
\end{figure}

The solution (4.52)-(4.54) of ref.\ \cite{Guralnik:2003we} has been reproduced in the framework of the first order approach to 2d gravity with the following generalizations: It is embedded into a larger patch of the geometry because the coordinate $X$ in (\ref{cs:6}) is not bounded by $\sqrt{c}$ as opposed to (4.52). Moreover, a second Casimir function is present and only for a special tuning between both Casimirs, $c^2=8\mathcal{C}^{(g)}$, the solution (4.54) is reproduced; otherwise, more general solutions emerge with up to 4 Killing horizons. Their global properties have been discussed. A summary of these results is contained in the ``phase-space'' plot fig.\ \ref{fig:1}.

A straightforward generalization of the formulation (\ref{cs:1}) would be the consideration of arbitrary $\mathcal{V}(X^+X^-,X,Y)$ instead of the special case (\ref{cs:2}). For all these models one Casimir function (corresponding to the total charge) becomes $Y=c$, while the other one is in general more complicated and related to the total energy. Possible applications of such models are twofold: if $Y$ appears at least quadratically in $\mathcal{V}$ it can be eliminated from the EOM obtained by varying with respect to $Y$ (not necessarily uniquely); in this case it represents the dual field strength (possibly with some coupling to the dilaton $X$). Such a situation is encountered e.g.\ for potentials of the type $\mathcal{V}=\tilde{\mathcal{V}}(X^+X^-,X)+F(X)Y^2$ including the spherically reduced Reissner-Nordstr\"om black hole. If, however, $Y$ appears only linearly in the form $\mathcal{V}=\tilde{\mathcal{V}}(X^+X^-,X)+F(X)Y$ as in the present case then the ``dilaton'' $X$ (or a function thereof) determines the dual field strength. This implies an interesting ``gauge curvature to geometric curvature'' coupling in the action which is explicit in the second order formulation (\ref{cs:0}). 

Further generalizations are conceivable, e.g.\ the coupling to matter fields thus making the theory nontopological. In that case the virtual black hole phenomenon should be present \cite{Grumiller:2000ah} and interesting results can be derived within the path integral formalism \cite{Kummer:1997hy}.

Indeed, powerful methods to study these models classically, semi-classically and at the quantum level already do exist \cite{Grumiller:2002nm}.

\section*{Acknowledgement}
 
This work has been supported by project P-14650-TPH of the Austrian Science Foundation (FWF). We are grateful to R.~Jackiw, T.~Strobl and D.~Vassilevich for helpful correspondence. DG renders special thanks to C.~B{\"o}hmer for support with {\tt xfig}.

%\bibliographystyle{../review01/fullsort} % for the author's convenience
%\bibliography{../review01/review} % my bibtex-file - just use the .bbl file instead or I can send you my bibtex file, if you want (but please do not include new refs in it - rather let me include them, it is much easier to handle this ``centralized''

\input{note.bbl.fix}

\end{document}

%% file: macros
\newcommand{\beqs}{\begin{equation*}}
\newcommand{\beq}{\begin{equation}}

\newcommand{\eeqs}{\end{equation*}}
\newcommand{\eeq}{\end{equation}}

\newcommand{\beqas}{\begin{eqnarray*}}
\newcommand{\beqa}{\begin{eqnarray}}

\newcommand{\eeqas}{\end{eqnarray*}}
\newcommand{\eeqa}{\end{eqnarray}}

\newcommand{\eq}[2]{\begin{equation} #1 \label{#2} \end{equation}}

\newcommand{\eps}{\varepsilon}
\newcommand{\al}{\alpha}
\newcommand{\be}{\beta}

\newcommand{\de}{\delta}
\newcommand{\om}{\omega}

\newcommand{\blist}{\begin{itemize}}

\newcommand{\elist}{\end{itemize}}

\providecommand{\href}[2]{#2}

\DeclareFontFamily{OT1}{rsfs}{}
\DeclareFontShape{OT1}{rsfs}{m}{n}{ <-7> rsfs5 <7-10> rsfs7 <10->rsfs10}{} 
\DeclareMathAlphabet{\mycal}{OT1}{rsfs}{m}{n}

%% file: CS.title.new
\begin{titlepage}

\renewcommand{\thefootnote}{\fnsymbol{footnote}}

\hfill TUW--03--18

% \hfill Vers. 1.0 -- DRAFT \today\\ %% just for draft versions 

\begin{center}
\vspace{0.5cm}

{\Large\bf The Classical Solutions of the Dimensionally Reduced Gravitational Chern-Simons Theory}
\vspace{1.0cm}
% \vfill

{\bf D.\ Grumiller\footnotemark[1], W.\ Kummer\footnotemark[2] % and D.V.\ Vassilevich\footnotemark[3]\footnotemark[4]
}
\vspace{7ex}

  {\footnotemark[1] \footnotemark[2]
\footnotesize Institut f\"ur
    Theoretische Physik, Technische Universit\"at Wien \\ Wiedner
    Hauptstr.  8--10, A-1040 Wien, Austria}
\vspace{2ex}

  %{\footnotemark[3]\footnotesize Intitut f\"{u}r Theoretische Physik, Universit\"{a}t Leipzig,\\Augustusplatz 10, D-04109 Leipzig, Germany}

   \footnotetext[1]{e-mail: \texttt{grumil@hep.itp.tuwien.ac.at}}
   \footnotetext[2]{e-mail: \texttt{wkummer@tph.tuwien.ac.at}}
   %\footnotetext[3]{e-mail: \texttt{vassil@itp.uni-leipzig.de}}
   %\footnotetext[4]{On leave from V.Fock Institute of Physics,St.Petersburg University, 198904 St.Petersburg, Russia}
\end{center}
\vspace{7ex}

\begin{abstract}

%%% ABSTRACT STARTS HERE %%%

The Kaluza-Klein reduction of the 3d gravitational Chern-Simons term to a 2d theory is
equivalent to a Poisson-sigma model with fourdimensional target space
and degenerate Poisson tensor of rank 2. Thus two constants of motion
(Casimir functions) exist, namely charge and energy. The
application of well-known methods developed in the framework of first
order gravity allows to construct all classical solutions
straightforwardly and to discuss their global structure. For a certain
fine tuning of the values of the constants of motion the solutions of
\href{http://www.arXiv.org/abs/hep-th/0305117}{{\tt hep-th/0305117}}
are reproduced. Possible generalizations are pointed out.

%%% END OF ABSTRACT %%%

\end{abstract}

%PACS numbers: 
% 04.60.Kz; 04.60.Gw; 11.10.Lm; 11.80.Et

\vfill
\end{titlepage}

%% file: note.bbl.fix
\providecommand{\href}[2]{#2}\begingroup\raggedright\endgroup